# Indirect-to-direct band-gap crossover in few-layer MoTe$_2$

*Ignacio Gutiérrez Lezama,\*[†,‡] Ashish Arora,[§] Alberto Ubaldini,[†,¥] Céline Barreteau,[†] Enrico Giannini,[†] Marek Potemski[§] and Alberto F. Morpurgo\*[†,‡]*

[†]DPMC and [‡]GAP, Université de Genève, 24 quai Ernest Ansermet, CH-1211 Geneva, Switzerland

[§]Laboratoire National des Champs Magnétiques Intenses (LCNMI), CNRS, 25 rue des Martyrs B.P. 166, 38042 Grenoble, France

ABSTRACT: We study the evolution of the band-gap structure in few-layer MoTe$_2$ crystals, by means of low-temperature micro-reflectance (MR) and temperature-dependent photoluminescence (PL) measurements. The analysis of the measurements indicate that –in complete analogy with other semiconducting transition metal dichalchogenides (TMDs)– the dominant PL emission peaks originate from direct transitions associated to recombination of excitons and trions. When we follow the evolution of the PL intensity as a function of layer thickness, however, we observe that MoTe$_2$ behaves differently from other semiconducting TMDs investigated earlier. Specifically, the exciton PL yield (integrated PL intensity) is identical for mono and bilayer and it starts decreasing for trilayers. A quantitative analysis of this behavior and of all our experimental observations is fully consistent with mono and bilayer MoTe$_2$ being direct band-gap semiconductors, with tetralayer MoTe$_2$ being an indirect gap semiconductor, and with trilayers having nearly identical direct and indirect gaps. This conclusion is different from the one reached for other recently investigated semiconducting transition metal dichalcogenides, for which only monolayers are found to be direct band-gap semiconductors, with thicker layers having indirect band gaps that are significantly smaller –by hundreds of meV- than the direct gap. We discuss the relevance of our findings for experiments of fundamental interest and possible future device applications.

Thin layers of semiconducting transition metal dichalcogenides (TMDs) have received much attention due to their very interesting and rich opto-electronic properties,[1–4] which largely originate from the indirect-to-direct band gap crossover occurring when these materials are thinned down to a single monolayer (ML).[5–11] This crossover is not only responsible for a large enhancement of the photoluminescence (PL) quantum yield,[5,7] resulting in a strong increase in the intensity of emitted light, but is also an essential condition for the observation of exciting phenomena resulting from valley-selective optical transitions,[12–17] such as the valley-Hall effect[17,18] and the possibility to control the spin and valley degrees of freedom.[13–16] First observed in $MoS_2$,[5,6] the indirect-to-direct band gap crossover in monolayers has been found to be a common property of all the semiconducting TMDs investigated so far, including $WS_2$,[7] $WSe_2$[7] and $MoSe_2$[8,9]: the only difference between all these materials is of quantitative nature, i.e., the magnitude of the band gap, of the exciton binding energy, spin-orbit splitting, etc. The occurrence of such an identical behavior, with only quantitative differences in the material parameters, is being taken advantage of in the design of materials with engineered band structure, which can be achieved by alloying transition metals and chalcogens with appropriate stoichiometry.[19–22]

Recently we have started exploring (2H) $MoTe_2$, one of the TMDs that has received limited attention so far,[23–27] focusing on the structure of the band gap in thick (> 30 nm) exfoliated crystals[26]. We found that in these "bulk-like" crystals the difference between the indirect (0.88 eV) and direct band gap (1.02 eV) is small, approximately only 0.15 eV (four-to-five times smaller than in all other TMDs investigated so far; in $MoS_2$, for instance, this difference is approximately 0.7 eV[28,29]), and we suggested that –because of this small difference– the indirect-to-direct band gap crossover in $MoTe_2$ may occur already before reaching monolayer thickness. Here, we address this issue through a systematic study of the low-temperature PL and reflectivity in $MoTe_2$ crystals ranging from 1 to 5 MLs. We find evidence that the crossover from indirect to direct band gap occurs before the monolayer limit, differently from the conclusion reported in a very recent study analogous to ours (see discussion at the end) [27]. We reach this conclusion by characterizing the direct excitonic transition observed in PL and micro-reflectance (MR) measurements, and by analyzing the evolution of the relative exciton PL yield (integrated PL intensity relative to the monolayer) with increasing thickness. We find that the behavior of mono and bilayer is fully consistent with that of a direct band-gap semiconductor, that tetralayers are indirect band-gap semiconductors, and that trilayers have nearly identical direct and indirect gaps.

The atomically thin MoTe$_2$ layers studied here were exfoliated from both commercial (2Dsemiconductors Inc.) and in-house grown bulk crystals,[30] with no significant difference in their optical properties. The exfoliated MoTe$_2$ layers were transferred onto Si substrates covered with a 290 nm SiO$_2$ layer, and their thickness was determined through combined atomic force microscopy (AFM; Figure 1a) imaging, Raman spectroscopy (Figure 1b and 1c) and optical contrast measurements (Figure 1d). Specifically, the characterization of over 30 samples with thicknesses ranging from 1 to 6 MLs allowed us to identify two methods that can be used to determine the thickness unambiguously, in a fast and non-invasive manner. These are the evolution of the relative intensity of Raman modes (Figure 1c), and the evolution of the difference in optical contrast with the underlying substrate (Figure 1d). Starting with the discussion of the Raman spectra, Fig. 1b shows data for few-layer MoTe$_2$ crystals, in which three distinctive peaks are observed, [23] the in-plane $E^1_{2g}$ mode (~232 cm$^{-1}$), the out-of plane $A_{1g}$ mode (~171 cm$^{-1}$) and the bulk in-active $B^1_{2g}$ phonon mode (~288 cm$^{-1}$, this latter peak is entirely absent in monolayers, which easily enables their identification). The positions of the $A_{1g}$ and $E^1_{2g}$ peaks follow a systematic shift with decreasing the number of layers (Figure 1b). Although, in principle, this shift can be used to determine the layer thickness, in practice, its magnitude is too small –and susceptible to sample to sample variations– to provide a reliable determination. Through a careful analysis, however, we succeeded in establishing that the relative peak intensities associated to the $A_{1g}$, $B^1_{2g}$ and $E^1_{2g}$ phonon modes do vary significantly and systematically upon varying the number of layers. As it is clear from Fig. 1c, the relative changes in intensity are sufficiently large to provide a reliable thickness determination. Additionally, Fig. 1d illustrates how the relative optical contrast $C$ between the MoTe$_2$ crystals and the substrate ($C = (I_{sub} - I_{flake})/I_{sub}$, where $I_{sub}$ and $I_{flake}$ are the intensity of the substrate and of the crystal, measured with a digital camera under an optical microscope) also evolves systematically with layer thickness. Looking at this contrast, therefore, also provides an unambiguous thickness determination.[26]

We have performed temperature dependent PL and low-temperature MR measurements on MoTe$_2$ crystals with thickness ranging from one to five layers. The PL measurements were performed using an Ar ion laser with an excitation wavelength of 488 nm and a spot size of ~2 µm. The PL emission from the sample was then dispersed using a 0.3 m focal length monochromator and detected with a liquid-nitrogen cooled InGaAs array detector. During the experiments, the laser power was varied between 0.5 and 128 µW, resulting in a linear response of the PL intensity. In turn the MR measurements were performed using collimated light from a 100 W tungsten halogen lamp focused through a pinhole of 150 µm diameter, which was then collimated again before it was focused on the sample (spot size 3-4 µm). The detection arrangement was similar to the one used during the PL measurements. During the measurements the samples were kept cold by direct contact with the cold finger of a liquid-He continuous-flow cryostat, which allowed us to vary the sample temperature between 4 and 300 K.

As a first step, we identify the origin of spectral features observed in the PL and MR spectra of atomically thin MoTe$_2$ layers. The temperature dependence of the normalized PL spectra of mono and bilayer are shown in Fig. 2a and 2b, respectively. The same general trend is observed in both cases. At low temperature (4.5 to 100 K) the PL line shape consists of a split-peak that evolves into a single peak as the spectral weight of the low-energy peak is transferred to the high-energy one with increasing temperature (at the lowest temperatures, the high-energy peak in the monolayer is less pronounced, and becomes more easily visible starting from 50-60 K). The split peak also undergoes a red shift with increasing temperature, leaving the energy difference between the two maxima constant (see Figure 2c and 2d). The above trends are all consistent with the expected temperature dependence of the PL emitted by 2D excitons (high-energy peak) and trions (low-energy peak), as it has been found previously in monolayer $MoSe_2$.[31] These observations therefore establish the origin of the optical transitions that we observe in $MoTe_2$. From the energy difference between the two peaks, we can also extract the trion binding energy[31] (see Fig. 2c and 2d), corresponding to 25 ± 1 meV and 17 ± 1 meV for monolayers and bilayers, respectively. Since as the temperature rises electrons (or holes) escape their bound trion state due to thermal fluctuations, the lower binding energies found in bilayers, as compared to monolayers, is consistent with the disappearance of the trion peak at lower temperatures (seen Fig. 2a and 2b). The value found for the trion binding energies are comparable to those reported for other ultrathin semiconducting TMDs, such as monolayer $MoSe_2$ (30 meV),[31] monolayer $MoS_2$ (18 meV),[32] and bilayer $WSe_2$ (30 meV),[33] demonstrating the consistency of our interpretation.

To confirm that the PL emission originating from excitons and trions is due to direct optical transitions, as expected, we have compared the PL spectra of few-layer $MoTe_2$ crystals to the low-temperature MR spectra (Figure 3a), which, on atomically thin layers, are only sensitive to direct optical transitions.[6,7] Figure 3a shows the low-temperature reflectance spectra in terms of the reflectivity contrast $R_C = (R_{C+S} - R_S)/(R_{C+S} + R_S)$, where $R_{C+S}$ and $R_S$ are the reflectance measured on the crystal and on the substrate. We find that all the spectra in Fig. 3a show a dip in reflectivity contrast at energies close to those of the exciton transition energies measured in layers of the same thickness (see Fig. 2a, 2b, and 4a; on the substrate used for MR measurements no trilayers were found, which is why the corresponding spectrum is not present in Fig. 3a). To compare quantitatively the position of the different features, we have extracted the precise value of the transition energy by fitting the $R_C$ spectra, using a standard multi-layer transfer matrix method,[34,35] in which the excitonic contribution to the dielectric function of the ultrathin $MoTe_2$ crystals was assumed to be a Lorentzian. Although the fits do not reproduce the entire background in the $R_C$ spectra, they do reproduce the transition region sufficiently reliably to enable the determination of the transition energies (see Figure 3a). The quantitative comparison between the transition energies extracted from the $R_C$ spectra and the exciton transition energies measured in the PL spectra is shown in Figure 4a (the PL spectrum of the pentalayer is not shown because the signal is only slightly larger than the background emission coming from the Si substrate, and cannot be disentangled from it). The good correspondence between PL and MR data (in all cases where the comparison could be made) is clear. It confirms that the observed transitions are direct, and consistent with what is expected from the so-called A-exciton, whose spectral weight is divided into an exciton and a trion contribution.[31]

Having identified the origin of the PL in few-layer $MoTe_2$ is due to direct transitions associated to the A exciton, we analyze the evolution of the transition energy and PL intensity with layer thickness. As shown in Figure 4a, the A exciton transition energy blue-shifts with decreasing thickness, while the associated PL intensity increases (all the spectra were measured at the same laser power). Although these trends have been observed in the previously investigated thin-layer semiconducting TMDs,[5,7,11] the behavior of MoTe$_2$ is distinctively different: when passing from bi to monolayer the increase of the maximum PL intensity is only a factor of 2-3 (see Figure 4a,b). This is much smaller than what is observed in other semiconducting TMDs, where the indirect-to-direct band gap crossover causes an increase in PL intensity that is between one and two orders of magnitude.[5,8] To ensure that what we observe is not an artifact of a specific sample in which the monolayer PL is coincidentally quenched by some extrinsic mechanism, we have performed measurements on many different MoTe$_2$ bi and monolayers and found a very high reproducibility of our observations, as shown in Fig. 4c. We emphasize that the observed behavior also demonstrates that the PL intensity is not affected by the unintentional presence of doping, which –as we have described in our previous study[26]– exhibits large random sample-to-sample fluctuations in concentration that do not correlate with the high reproducibility of the PL intensity.[36]

Another difference between MoTe$_2$ and previously studied thin-layer semiconducting TMDs is the absence of a PL peak associated to indirect transitions (Fig. 4a).[5,7] What is seen in other semiconducting TMDs is that monolayers –because of their direct-gap character– do not show a PL peak associated to indirect transitions at higher energy (since electron-hole pairs relax rapidly to the K point in the Brillouin zone, form an exciton, and recombine undergoing a direct transition). In bilayers, however, the indirect gap is significantly smaller than the direct one, and a large number of electron-hole pairs can relax in k-space to form indirect excitons (i.e., excitons with electrons and holes having different momenta). Part of these excitons decay radiatively leading to a measurable PL, whose intensity is experimentally found to be smaller than that of the direct exciton transition (measured in the same bilayer) by at most a factor of 3-4.[5,7]

It is clear from the data of Fig. 4a that in our experiments, an indirect transition with the intensity expected based on the analogy with other semiconducting TMDs would be well above the background, and therefore it should be detected if present. However, in the case of $MoTe_2$ no transition is seen experimentally. The absence of an indirect transition, as well as the small difference in PL intensity between the direct transitions seen in mono and bilayers, suggest that mono and bilayers of $MoTe_2$ are direct gap semiconductors, as we had originally hypothesized. Support for this conclusion is drawn from the quantitative analysis of the integrated PL intensity of each layer. The relative exciton PL yield $I/I_{mono}$, where $I$ is the integrated exciton PL intensity (trion plus exciton spectral weight) of each layer and $I_{mono}$ is that of the monolayer, is plotted in Fig. 4d for crystals ranging from 1 to 4 MLs. Rather strikingly, for $MoTe_2$ the PL yield of bilayers coincides with that of monolayers. This is incompatible with bilayers having an indirect transition at energy smaller than, or coincident with, that of the direct transition. In turn, the PL yield of $MoTe_2$ trilayers is smaller than that of monolayers, but only by a factor between 3 and 4, indicating that the direct and indirect transitions have approximately the same energy, i.e., the direct and indirect gap approximately coincide (and it is difficult to establish conclusively from this data whether one is larger than the other). This is likely also the reason why trilayers do not show an indirect transition: such a transition occurs at essentially the same energy as the direct one, but has a much smaller intensity, and its experimental manifestation is therefore eclipsed (i.e., in trilayers the indirect transition is masked by the direct one, which is more intense and virtually at the same energy). Finally, $MoTe_2$ tetralayers exhibit 25-30 times smaller PL yield than monolayers (possibly even more because –in the case of tetralayers– the background does give a non-negligible contribution to the value of $I$ extracted from the data), which establishes unambiguously their indirect-gap semiconductor nature (note that in tetralayers the intensity of the indirect transition would not be much larger than the background, and hence it would be difficult to detect). We conclude that a scenario in which mono and bilayer $MoTe_2$ are direct band gap semiconductors is consistent with all our experimental observations and that no alternative possibilities to account for all our findings can be easily conceived.

Our conclusion is different from that of Ruppert et al.,[27] who recently reported experiments analogous to ours (but mainly focusing on room-temperature measurements), and inferred from their measurements that bilayer $MoTe_2$ is an indirect band gap semiconductor. A key difference between their and our experimental observations is the relation between the energy of the transitions observed in the PL and MR measurements. Ruppert et al.[27] observe that, except for monolayers, the PL signal always occurs at an experimentally significantly lower energy than the MR signal (whereas in our case the two energies coincide, see Fig. 3b). Since in very thin layers a MR signal is only seen for direct transitions, they are forced to conclude that two transitions are actually present, and that the one at lower energy, manifesting itself in the PL signal and not in the MR measurements, must be an indirect transition (which implies that bi and thicker layers are indirect gap semiconductors). In this scenario, however, it is difficult to understand why the direct transition inferred from MR measurements does not give any visible PL signal. As explicitly discussed in Ref. 27, the high temperature at which their measurements are carried out complicates the interpretation, calling for systematic low temperature measurements. These low temperature measurements are the focus of the work presented here, and give experimental results that lead to the scenario explained above, which leads us to conclude that both mono and bilayers are direct band gap semiconductors.

We conclude that when the thickness of the material is reduced to a few atomic layers, the behavior of MoTe$_2$ differs somewhat from that of previously studied semiconducting TMDs. This finding is interesting: whereas previous studies appeared to establish the notion that all few-layer semiconducting TMDs have a qualitatively identical behavior, differing only for the precise values of the relevant energy scales, our results show that care needs to be taken. For fundamental physics, finding that MoTe$_2$ bilayers have a high PL intensity can have implications for interesting future experiments. For instance, recent theoretical calculations predict the closing of the band gap (at the K-K points) in bi-layer semiconducting TMDs upon the application of a perpendicular electric field,[37,38] and in MoTe$_2$ bilayers the enhanced PL could provide a particularly sensitive way to probe the evolution of the gap experimentally (e.g., in devices with ionic liquid gates that enable large perpendicular electric fields to be applied). In terms of potential for electronic applications, our results are also of interest. As it is common to all semiconducting TMDs investigated so far, a large PL combined with balanced (and sufficiently high) electron and hole mobility values (which for MoTe$_2$ has already been reported in thicker exfoliated crystals[26]) are potentially useful for the realization of light-emitting devices. In this context, mono and bilayer MoTe$_2$ are particularly interesting, because having excitons at much lower energies than other few layers TMDs, they can enable the realization of devices operating in the far infrared. Finally, we also note that the excitonic transitions in MoTe$_2$ mono and bilayers are very close to the 1.3 eV band gap needed for maximum efficiency in single junction solar cells,[3] making these system worth investigating for photovoltaic applications.

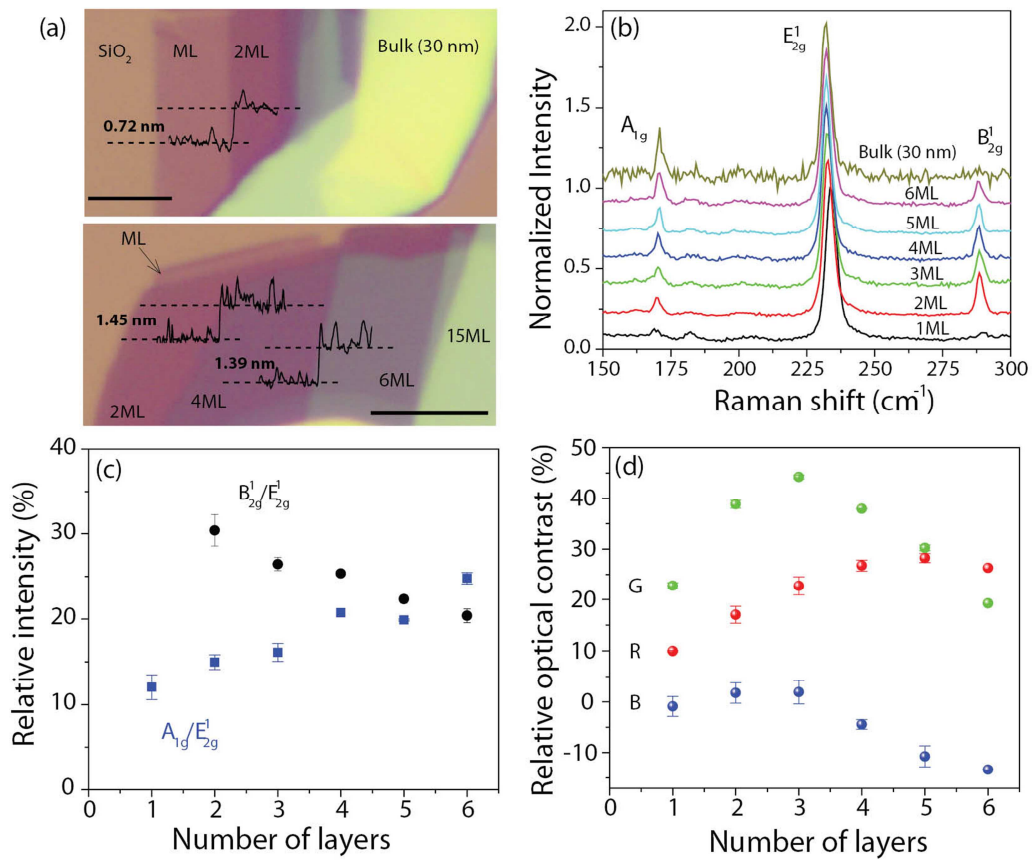

**Figure 1.** Characterization and identification of few-layer MoTe$_2$. (a) Optical microscope images of multi-layer MoTe$_2$ crystals and their corresponding AFM step height. The scale bars are 5 μm. (b) Raman spectra of a bulk and ultrathin MoTe$_2$ crystals with thicknesses ranging from 1 to 6 monolayers (ML). Curves offset for clarity. (c) Evolution of the relative Raman intensity of the $A_{1g}$ and $B^1_{2g}$ phonon modes with respect to the $E^1_{2g}$ mode. (d) Evolution of the relative optical contrast between the MoTe$_2$ crystals and the 290 nm SiO$_2$ substrate, in the red (R), green (G) and blue (B) channels.

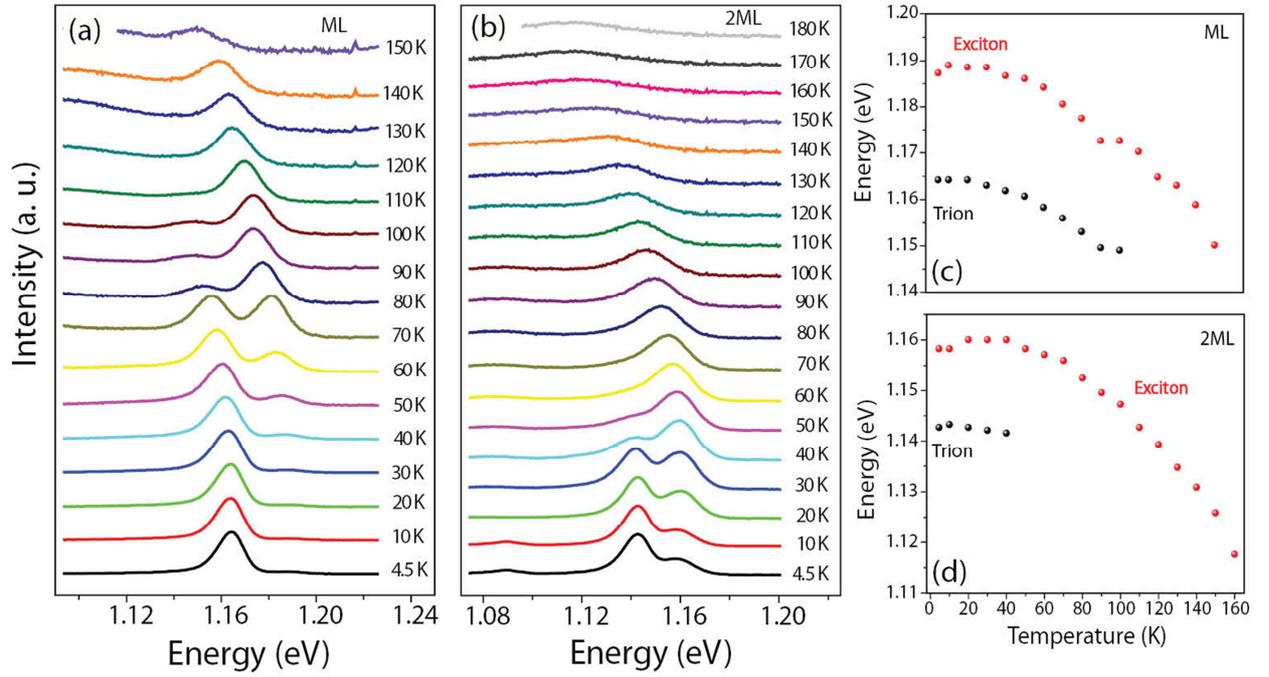

**Figure 2.** Normalized temperature-dependent PL spectra of mono-(a) and bi-layer (b) MoTe$_2$ measured with a 125 and 50 μW laser power, respectively. The observed PL is by the A exciton, in which at low temperature the exciton and trion contributions are resolved. The curves are offset for clarity. (c) mono and (d) bilayer exciton and trion transition energies extracted from the spectra shown in (a) and (b). The difference between the two excitations is constant and corresponds to the trion binding energy, as discussed in the main text.

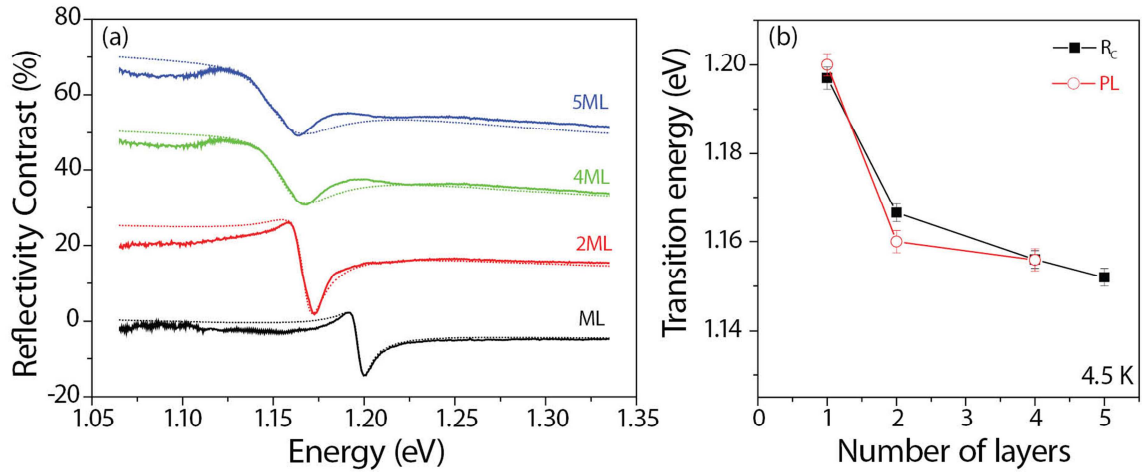

**Figure 3**. (a) Reflectivity contrast spectra of few-layer $MoTe_2$ measured at T = 4.5 K. The spectra, which have been offset for clarity, show a dip corresponding to the A exciton transition, whose energy undergoes a blue shift with decreasing layer thickness. The dotted lines are the theoretical fits to extract the transition energies, as discussed in the main text. Panel (b) shows that the position of the transition energy extracted from the reflectivity measurements (full squares) and from PL (open circles) are in very good agreement.

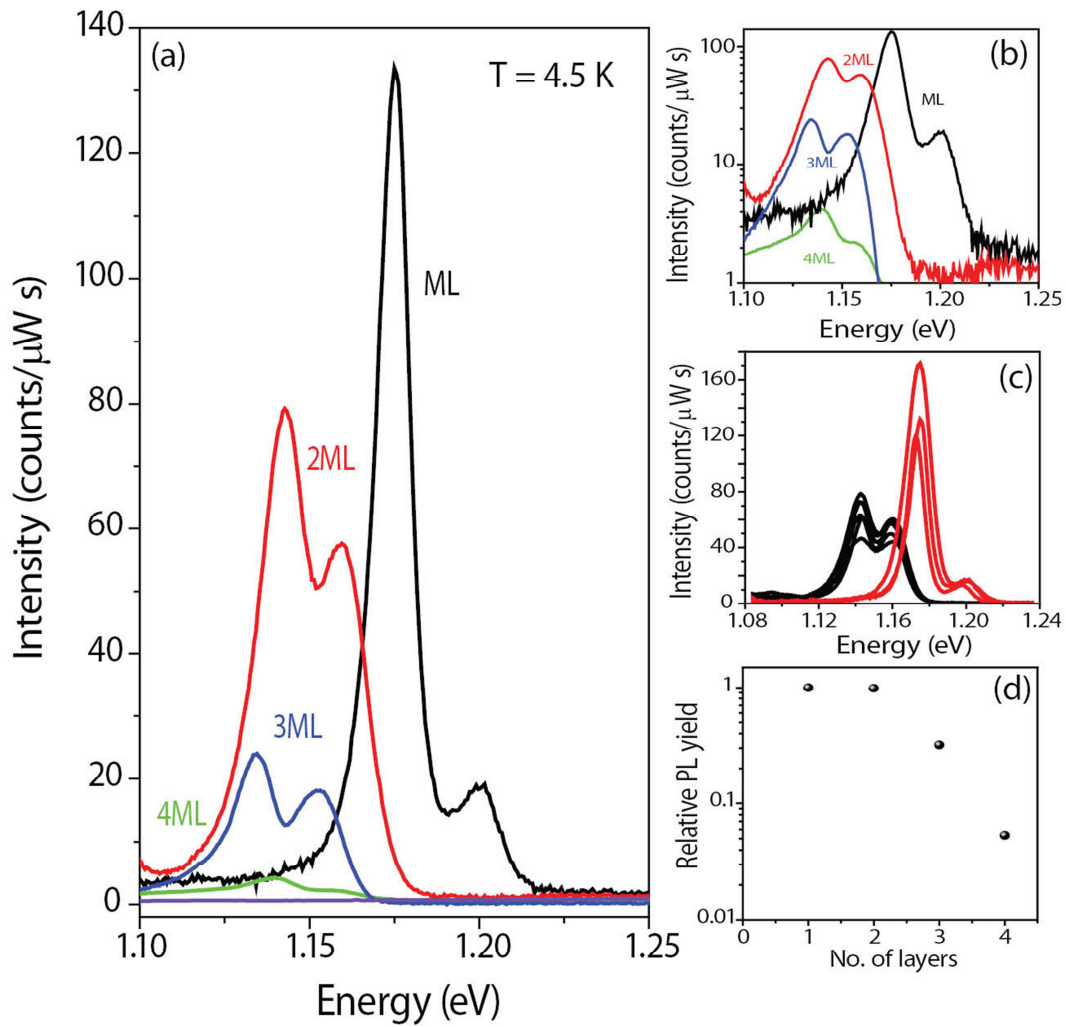

**Figure 4.** Photoluminescence spectra of few-layer MoTe$_2$ measured at T = 4.5 K and a laser power of 128 μW, plotted in linear (a) and semi-log scale (b). The two peaks observed in the spectra originate from the radiative recombination of trions (low-energy peaks) and excitons (high-energy peaks), whose transition energies undergo a blue shift with decreasing layer thickness. Panel (c) illustrates the sample-to-sample reproducibility of the results, by showing PL spectra measured on several MoTe$_2$ mono and bilayers. Panel (d) shows the evolution of the relative exciton PL yield for MoTe$_2$ layers of different thickness (from one to four layers).

AUTHOR INFORMATION

**Corresponding Author**

*Email: Ignacio.gutierrez@unige.ch (I.G.L.);  alberto.morpurgo@unige.ch (A.F.M.).

**Present Addresses**

¥Università degli Studi di Salerno, Istituto Superconduttori, Materialiari innovativi e Dispositivi del Consiglio Nazionale delle Ricerche, Via Giovanni Paolo II, 132-84084 – Fisciano (SA), Italia.

**Author Contributions**

I.G.L. prepared and identified the MoTe$_2$ layers and did the optical measurements together with A.A. A.U., C.B., and E.G. grew and characterized the MoTe$_2$ crystals used to produce ultrathin crystalline layers. I.G.L., A.A., M.P., and A.F.M. analyzed and interpreted the data. I.G.L. did most of the manuscript writing, with significant input from A.A., M.P., and A.F.M.. All authors discussed the results and commented the final manuscript.

**Notes**

The authors declare no competing financial interest.

ACKNOWLEDGMENT

We gratefully acknowledge A. Ferreira for technical support, A. Kuzmenko for fruitful discussions, N. Ubrig for support with the Raman measurements and S. Ghosh for useful suggestions during the building of the micro-reflectance setup. Financial support from the Swiss National Science Foundation, the European Graphene Flagship and the ERC-2012-AdG-320590-MOMB grant is also acknowledged.